\documentclass[twoside]{ilcws07}
\usepackage[latin1]{inputenc}
\usepackage[dvips]{graphicx,epsfig,color}
\usepackage{wrapfig,rotating}
\usepackage{amssymb,amsmath,array}

\begin{document}
\title{
Off-Shell and Interference Effects for SUSY Particle Production} 
\author{J\"urgen Reuter
\vspace{.3cm}\\
Albert-Ludwigs-Universit\"at Freiburg -  Physikalisches Institut \\
Hermann-Herder-Str.~3, D-79104 Freiburg - Germany
}

\maketitle

\begin{abstract}
  We show that the narrow-width approximation is insufficient for
  describing production of supersymmetric particles at the
  ILC. Especially when cuts are taken into account to extract signals 
  using the narrow-width approximation can be wrong by an order of
  magnitude. 
\end{abstract} 
 
 \section{Precision SUSY measurements}

\begin{figure}
  \begin{center}
    \includegraphics[width=.75\textwidth,height=6.2cm]
		    {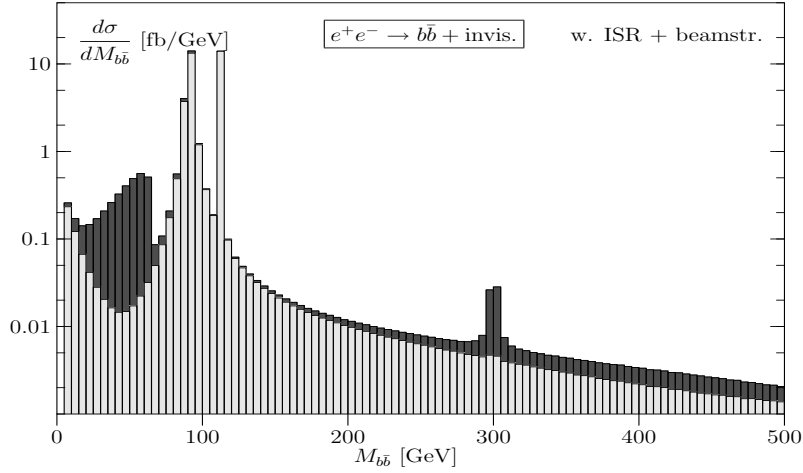} 
  \end{center}
  \caption{The $b\bar{b}$ invariant mass spectrum for the full process
  $e^+e^-\to b\bar{b}$+$E\hspace{-0.6em}/\hspace{0.2em}$ with ISR and
  beamstrahlung.  The SM background ($Z\to\nu\bar{\nu}$) with the
  $Z,h$ peaks is light gray. Dark gray represents all MSSM processes,
  with two peaks from heavy neutralino and heavy Higgs decays.}
  \label{fig:fullspectrum}
\end{figure}

 Supersymmetry (SUSY) is the best-motivated solution to the hierarchy
 problem. If SUSY is realized in Nature, the LHC is likely to find
 sparticles during the next couple of years. The precision
 spectroscopy of the new particles will then be the major goal of
 future particle physics experiments. The aim is to perform mass
 measurements to get the spectrum (edges in decay chains), to access
 the spin of all new particles via angular/spin correlations, and
 finally to perform coupling measurements to verify SUSY by the
 relations among the couplings. Therefore, we need precise predictions
 SUSY processes: for their own determination as well as because they
 are background for (more difficult) SUSY processes. We need parameter
 values as precise as possible in order to reverse the
 renormalization-group evolution and get a handle on the GUT
 parameters~\cite{pmz,spa}.   
 Corrections to (SUSY) processes (at the ILC) can be grouped into six
 categories~\cite{catpiss}: 1) Loop corrections to SUSY production and
 decay processes; 2) nonfactorizable, maximally resonant photon
 exchange between production and decay; 3) real radiation of
 photons/gluons; 4) off-shell kinematics for the signal process (see
 also~\cite{nonwa}); 5) irreducible background from all other SUSY
 processes; 6) reducible, experimentally indistinguishable SM
 background processes. Topics 1) and 3) are addressed in~\cite{susynlo}.
 

\section{Complexity and Approximations}

 Generic SUSY processes have an incredible complexity:
 e.g.~$e^+e^- \to b\bar b e^+e^- \tilde{\chi}^0_1\tilde{\chi}^0_1$
 (which is just an 
 exclusive final state for $\tilde{\chi}^0_2\tilde{\chi}^0_2$
 production) has 66,478 
 diagrams already at tree level. Entangled in these amplitudes are 
 different signal diagrams: $e^+e^- \to \tilde{\chi}^0_i
 \tilde{\chi}^0_j, \tilde{b}_i \tilde{b}_j, \tilde{e}_i \tilde{e}_j$. 
 To disentangle them in simulations or a real data analysis, one has
 to use cuts and to consider SM backgrounds (here e.g.~$e^+e^- \to
 b\bar b e^+ e^- \nu_i\bar\nu_i$). There are much more complicated
 processes for LHC, and even for ILC. To deal with this complexity one
 needs to use multi-particle event generators~\cite{snowmass}.  

 There are three different levels of approximations used for
 describing such processes like $A_1 A_2 \to P^{(*)} \to F_1 F_2$: the
 narrow-width approximation $\sigma(A_1 A_2 \to P) \times {\rm BR}(P
 \to F_1 F_2)$ (on-shell production times branching ratio), the
 Breit-Wigner approximation $\sigma(A_1 A_2 \to P) \times \frac{M_P^2
 \Gamma_P^2}{(s-M_P^2)^2 + \Gamma_P^2 M_P^2} \times {\rm BR}(P \to F_1
 F_2)$ (folding in a finite width propagator), and the full matrix
 elements: $\sigma(A_1 A_2 \to F_1 F_2)$. That last level is not
 featured by event generators like ISAJET, PYTHIA, HERWIG, SUSYGEN.

\begin{table}
 \begin{center}
   \begin{small}
   \begin{displaymath}
     \begin{array}{|l|rrr|}
       \hline
       \text{Channel} & \sigma_{2\to 2} & 
       \sigma\times\mathrm{BR} & \sigma_{\rm BW}
       \\
       \hline
       Zh& 20.574 & 1.342 &
       1.335 
       \\
       ZH           &  0.003 & 0.000 & 0.000
       \\
       HA           &  5.653 &
       0.320 & 0.314
       \\
       \tilde{\chi}^0_1\tilde{\chi}^0_2   & 69.109 &
       13.078 & 13.954
       \\
       \tilde{\chi}^0_1\tilde{\chi}^0_3   & 24.268 &
       3.675 & 4.828 
       \\
       \tilde{\chi}^0_1\tilde{\chi}^0_4   & 19.337 & 0.061 & 0.938
       \\
       \tilde{b}_1\tilde{b}_1     &  4.209 &
       0.759 & 0.757 
       \\
       \tilde{b}_1\tilde{b}_2     &  0.057 &
       0.002 & 0.002 
       \\
       \hline
       \text{Sum}   &        &19.238 &22.129
       \\
       \hline\hline
       \text{Exact} &        &       &19.624
       \\
       \text{w/ISR}         &&       &22.552
       \\
       \hline
       \hline\hline
       Z\bar\nu\nu  & 626.1 & 109.9 & 111.4
       \\
       h\bar\nu\nu  & 170.5 &  76.5 &  76.4
       \\
       H\bar\nu\nu  &   0.0 &   0.0 &   0.0
       \\
       \hline
       \text{Sum}   &       & 186.5 & 187.7
       \\
       \hline\hline
       \text{Exact} &       &       & 190.1
       \\
       \text{w/ISR} &       &       & 174.2
       \\
       \hline
     \end{array}
   \hspace*{9mm}
     \begin{array}{|l|rr|}
       \hline
       \text{Channel} &  \sigma_{\rm BW} & \sigma_{\rm
         BW}^{\rm cut}
       \\
       \hline
       Zh           & 1.335 & 0.009
       \\
       HA           & 0.314 & 0.003
       \\
       \tilde{\chi}^0_1\tilde{\chi}^0_2   &13.954 & 0.458
       \\
       \tilde{\chi}^0_1\tilde{\chi}^0_3   & 4.828 & 0.454
       \\
       \tilde{\chi}^0_1\tilde{\chi}^0_4   & 0.938 & 0.937
       \\
       \tilde{b}_1\tilde{b}_1     & 0.757 & 0.451
       \\
       \tilde{b}_1\tilde{b}_2     & 0.002 & 0.001
       \\
       \hline
       \text{Sum}   &22.129 & {\mathbf 2.314}
       \\
       \hline\hline
       \text{Exact} &19.624 & {\mathbf 0.487}
       \\
       \text{w/ISR} &22.552 & {\mathbf 0.375}
       \\
       \hline
       \hline\hline
       Z\bar\nu\nu  & 111.4 &  2.114
       \\
       h\bar\nu\nu  &  76.4 &  0.002
       \\
       H\bar\nu\nu  &   0.0 &  0.000
       \\            
       \hline
       \text{Sum}   & 187.7 &  2.117
       \\
       \hline\hline
       \text{Exact} & 190.1 &  1.765
       \\
       \text{w/ISR} & 174.2 &  1.609
       \\
       \hline
     \end{array}
   \end{displaymath} 
   \end{small}
 \end{center}
 \caption{Main subprocesses for sbottom production at an 800 GeV ILC
 using the three level of complexity mentioned in the text. Left:
 before the cuts, right: after the cuts. Upper table is signal
 processes, lower one SM backgrounds. All processes in femtobarn.} 
 \label{tab:sigbkgd_crosssec}
\end{table}

 The simulations presented here have been performed with the
 multi-purpose event generator \texttt{WHIZARD}~\cite{whizard}, which
 is well-suited for physics beyond the
 SM~\cite{omwhiz_bsm}. Especially, the MSSM implementation has been
 thoroughly tested~\cite{catpiss}, e.g.~in a comparison with the other
 two MSSM multi-particle generators, \texttt{Madgraph} and
 \texttt{Sherpa}. The reference data can be found at
 \texttt{http://whizard.event-generator.org/susy\_comparison.html}.  

 \begin{figure}
   \begin{center}
        \includegraphics[width=.75\textwidth,height=6.4cm]
			{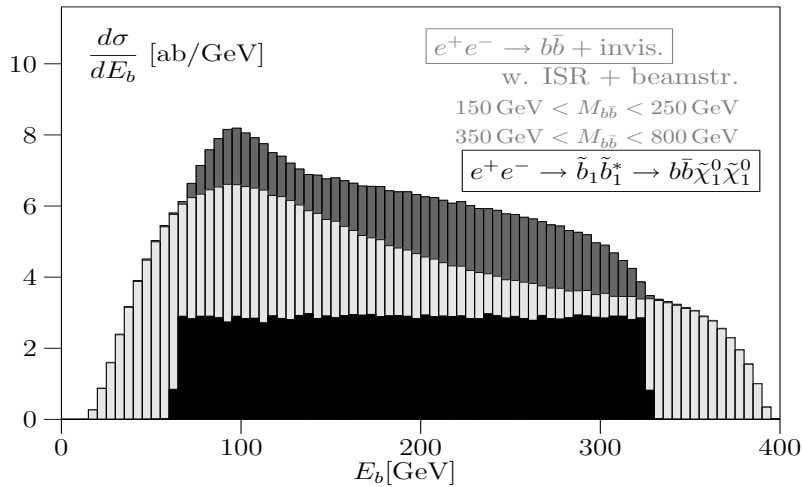}
   \end{center}
   \caption{The $E_b$ spectrum of $e^+e^-\to
  b\bar{b}$+$E\hspace{-0.6em}/\hspace{0.2em}$, including 
  all interferences and off-shell effects, plus ISR and beamstrahlung.
  The light gray histogram is the SM background, dark gray the sum of
  SUSY processes, including the cuts. We also show the idealized case
  (red) of on-shell sbottom production without ISR or beamstrahlung.}
   \label{fig:sbottom_box}
 \end{figure}


\section{Results}

 For our study~\cite{catpiss}  of off-shell and interference effects
 and to test the quality of the Breit-Wigner approximation, we took a
 SUGRA-inspired parameter point with non-universal right-handed scalar
 masses and $\tan\beta = 20$. Note, that the following does not depend
 on this special point, however. This point features a light Higgs,
 directly above LEP limit~\cite{higgsoverview}, large ($47$ \%)
 invisible Higgs decays to the LSP, $m_{\tilde{q}} \sim 430$ GeV,
 light sbottoms accessible at the ILC, and is compatible with all
 low-energy data: $b\to s\gamma$, $B_s \to\mu^+\mu^-$, $\Delta\rho$,
 $g_\mu - 2$, CDM. The sbottoms have masses of 295.36 and 399.92 GeV
 and widths of 0.5295 and 3.4956 GeV, respectively. The neutralino
 masses are 46.84, 112.41, 148.09 and 236.77 GeV, their widths 0,
 0.00005, 0.01162 and 1.0947 GeV, respectively. The focus lies on
 BR$(\tilde{b}_1 \to b \tilde{\chi}^0_1) = 43.2 \%$, as we want to
 study sbottom 
 production at an 800 GeV ILC. 

 In contrast to the LHC, at the ILC sbottoms are produced by
 electroweak interactions. Hence, much more channels contribute to the
 same exclusive final state, $e^+ e^- \to b \bar{b}
 \tilde{\chi}^0_1\tilde{\chi}^0_1$:  
 $e^+ e^- \to Zh, ZH, Ah, HA, \tilde{\chi}^0_1\tilde{\chi}^0_2,
 \tilde{\chi}^0_1\tilde{\chi}^0_3, \tilde{\chi}^0_1\tilde{\chi}^0_4, 
 \tilde{b}_1\tilde{b}_1^*, \tilde{b}_1\tilde{b}_2^*$, altogether 412
 diagrams. The irreducible SM 
 background is $e^+e^- \to b\bar b \nu_i \bar{\nu}_i$ ($WW$ fusion,
 $Zh$, $ZZ$, 47 diagrams). Important is to use widths to the same
 order as your process, i.e.~tree level in our
 case. The left of Tab.~\ref{tab:sigbkgd_crosssec} shows the cross
 sections of the contributing subprocesses in the three levels of
 complexity described above. The $b \bar b$ invariant mass spectrum
 (dark) including the SM background (light gray) is shown in
 Fig.~\ref{fig:fullspectrum}. Light gray peaks stem 
 from the $Z$ and light Higgs resonance, while the dark gray peak
 comes from the heavy Higgses. The broad dark continuum at low
 energies results from heavy neutralinos. Hence, to isolate the SUSY 
 signal it is mandatory to cut out the resonances, namely the two
 windows $M_{b\bar b} < 150$ GeV and $250\,\text{GeV} < M_{b \bar b} <
 350$ GeV. The off-shell decay $\tilde{\chi}^0_3 \to
 (\tilde{b}_1)_{off} \bar b \to b 
 \bar b \tilde{\chi}^0_1$ gives a broad continuum instead of a
 well-defined peak 
 expected from subsequent 2-body decays; this causes some of the
 effects described below. ISR and beamstrahlung give corrections of
 the same order  as off-shell effects and affect all $p_{miss}$
 observables. The corresponding plots can be found in~\cite{catpiss}.
 The cross sections after application of the cuts are shown on the
 right of Tab.~\ref{tab:sigbkgd_crosssec}; note the difference between
 the exact result 0.487 fb and the Breit-Wigner approximation of 2.314
 fb showing a deviation of an order of magnitude. 
 Fig.~\ref{fig:sbottom_box} shows that the $\tilde{b}_1 \to b
 \tilde{\chi}^0_1$ decay 
 kinematics is affected by the off-shell and interference effects, the
 SM backgrounds as well as ISR and beamstrahlung in a way that makes
 it much harder to precisly extract the sbottom mass as desired.
       

In summary, precision predictions for SUSY phenomenology are
important, especially higher order virtual and real corrections. The 
factorization of processes into $2\to 2$ production and decay is
insufficient or even wrong. Off-shell effects and interferences affect
the results, especially with cuts. Therefore one has to use full
matrix elements~(cf.~\cite{joanne}), available from multi-particle
event generators where \texttt{WHIZARD} is especially well-suited for
ILC.  

 
\section{Acknowledgments} 
 
JR was partially supported by the Helmholtz-Gemeinschaft under
Grant No. VH-NG-005.

 
\section{Bibliography} 
 
\begin{footnotesize} 

\end{footnotesize}
 
 

\begin{thebibliography}{99} 
\bibitem{url} Slides: \\
\verb$http://ilcagenda.linearcollider.org/getFile.py/$ 

\verb$           access?contribId=59&sessionId=69&resId=0&materialId=slides&confId=1296$

\bibitem{pmz}
  P.~Zerwas, these proceedings. 

\bibitem{spa}
  \texttt{http://spa.desy.de/spa};   
  J.~A.~Aguilar-Saavedra {\it et al.},
  Eur.\ Phys.\ J.\  C {\bf 46}, 43 (2006).

\bibitem{catpiss}
  K.~Hagiwara {\it et al.},
  Phys.\ Rev.\  D {\bf 73}, 055005 (2006).

\bibitem{nonwa}
  D.~Berdine, N.~Kauer and D.~Rainwater,
  arXiv:hep-ph/0703058.

\bibitem{susynlo}
  T.~Robens, these proceedings; W.~Kilian, J.~Reuter and T.~Robens, 
  Eur.\ Phys.\ J.\  C {\bf 48}, 389 (2006);
  AIP Conf.\ Proc.\  {\bf 903}, 177 (2007)

\bibitem{snowmass}
  J.~Reuter {\it et al.},
  arXiv:hep-ph/0512012;

\bibitem{whizard}
  \texttt{http://whizard.event-generator.org}; 
  W.~Kilian, T.~Ohl, J.~Reuter, 
  to appear in Comput.~Phys.~Commun.; hep-ph/0708.4233;
  M.~Moretti, T.~Ohl, J.~Reuter,
  hep-ph/0102195;
  J.~Reuter,
  arXiv:hep-th/0212154.

\bibitem{omwhiz_bsm}
  T.~Ohl and J.~Reuter,
  Eur.\ Phys.\ J.\  C {\bf 30}, 525 (2003); 
  Phys.\ Rev.\  D {\bf 70}, 076007 (2004);
  J.~Reuter, these proceedings, arXiv: 0708.4241 [hep-ph];
  arXiv: 0708.4383 [hep-ph];
  W.~Kilian and J.~Reuter,
  Phys.\ Rev.\ D {\bf 70} (2004) 015004;
  W.~Kilian, D.~Rainwater and J.~Reuter,
  Phys.\ Rev.\ D {\bf 71}, 015008 (2005);
  hep-ph/0507081;
  Phys.\ Rev.\  D {\bf 74}, 095003 (2006).
  M.~Beyer {\em et al.},
  Eur.\ Phys.\ J.\  C {\bf 48}, 353 (2006);
  W.~Kilian and J.~Reuter,
  hep-ph/0507099.

\bibitem{higgsoverview}
  S.~Heinemeyer {\it et al.},
  hep-ph/0511332;
  S.~Kraml {\it et al.},
  arXiv:hep-ph/0608079.

\bibitem{joanne}
  J.~Hewett, these proceedings.
  
\end{thebibliography}
\end{document}